\documentclass[12pt,preprint]{aastex}

\shorttitle{Forward Shock Proper-Motions of Kepler's SNR}
\shortauthors{S. Katsuda, et al.}

\begin{document}

\title{Forward Shock Proper Motions of Kepler's Supernova Remnant}

\author{S. Katsuda\altaffilmark{1}, H. Tsunemi\altaffilmark{1},
  H. Uchida\altaffilmark{1}, and M. Kimura\altaffilmark{1}
}

\altaffiltext{1}{Department of Earth and Space Science, Graduate School
of Science, Osaka University,\\ 1-1 Machikaneyama, Toyonaka, Osaka,
560-0043, Japan; katsuda@ess.sci.osaka-u.ac.jp}


\begin{abstract}

The X-ray structure of Kepler's supernova remnant shows a rounded
shape delineated by forward shocks.  We measure proper-motions of the
forward shocks on overall rims of the remnant, by using archival {\it
  Chandra} data taken in two epochs with time difference of 6.09 yr.
The proper-motions of the forward shocks on the northern rim are
measured to be 0$^{\prime\prime}$.076 ($\pm$0$^{\prime\prime}$.032 $\pm$0$^{\prime\prime}$.016) -- 0$^{\prime\prime}$.11 ($\pm$0$^{\prime\prime}$.014
$\pm$0$^{\prime\prime}$.016) yr$^{-1}$, while those on the rest of the rims are
measured to
be 0$^{\prime\prime}$.15 ($\pm$0$^{\prime\prime}$.017
$\pm$0$^{\prime\prime}$.016) -- 0$^{\prime\prime}$.30
($\pm$0$^{\prime\prime}$.048 $\pm$0$^{\prime\prime}$.016) 
yr$^{-1}$, here the first-term errors are statistical uncertainties
and the second-term errors are systematic uncertainties.  Combining
the best-estimated shock velocity of 1660$\pm$120\,km\,sec$^{-1}$
measured for Balmer-dominated filaments in the northern and central
portions of the remnant (Sankrit et al.\ 2005) with the proper-motions
derived for the forward shocks on the northern rim, we estimate the
distance of 3.3$^{+1.6}_{-0.4}$\,kpc to the remnant.  We measure
the expansion indices, $m$, (defined as $R \varpropto t^m$) to be
0.47--0.82 for most of the rims.  These values are consistent
with those expected in Type-{\scshape I}a SN explosion models, in
which the ejecta and the circumstellar medium have power-law density
profiles whose indices are 5--7 and 0--2, respectively.  Also, we
shuold note
the slower expansion on the northern rim than that on the southern
rim.  This is likely caused by the inhomogeneous circumstellar medium;
the density of the circumstellar medium is higher in the north than
that in
the south of the remnant.  The newly estimated geometric center,
around which we believe the explosion point exists, is located at
$\sim$5$^{\prime\prime}$ offset in the north of the radio center.

\end{abstract}

\section{Introduction}

Kepler's supernova remnant (SNR; SN 1604) is one of the historical
SNRs.  The X-ray structure of the remnant shows a clear circular shape
with a radius of about 100$^{\prime\prime}$.  The distance estimations
to the remnant have been scattered in a range of 3--5\,kpc (e.g.,
Green 1984; Schaefer 1994).  Although the far side of 5\,kpc supported
by H {\scshape I} observations (Reynoso \& Goss 1999) has been adopted
in recent literature, optical studies combining proper-motions with 
shock velocities of Balmer-dominated filaments have preferred a near
side of 2.9$\pm$0.4\,kpc (Blair et al.\ 1991) or
3.9$^{+1.4}_{-0.9}$\,kpc (Sankrit et al.\ 2005).   

The SN type of Kepler's SNR is very interesting.  Based on
the light curve of the Kepler's SN, Baade (1943) classified this remnant
as a result of Type-{\scshape I}a SN.  {\it Ginga} spectrum from this 
remnant showed a strong Fe K line, supporting that the Kepler's SN was
a Type-{\scshape I}a SN (Hatsukade et al.\ 1990).  {\it ASCA} spectrum 
of this remnant showed strong K-shell lines from Si, S, and Fe,
and the measured relative abundances supported Type-{\scshape I}a
origin (Kinugasa \& Tsunemi 1999).  Recent deep {\it Chandra} 
observations detected strong emission lines from heavy elements almost
everywhere in the remnant, again supporting Type-{\scshape I}a
origin (Reynolds et al.\ 2007).  On the other hand, the density of the
ambient medium around the remnant was estimated to be so high (at
least 1\,cm$^{-3}$; 
e.g., Hughes \& Helfand 1985) at a large height of $\sim$470\,pc
(assuming the distance of 4\,kpc to the remnant) from the Galactic
plane.  The high density of the ambient medium was considered to be a
signature of circumstellar material (CSM) which was blown off from a
progenitor star as a stellar wind.  In addition to the existence of
the dense surroundings, Nitrogen-rich materials as a result of
CNO-processing in a massive progenitor star were detected in some
optical knots (van den Bergh \& Kamper 1977; Dennefeld 1982).  These
facts suggested that the Kepler's remnant was core-collapse in origin.
Also, Bandiera (1987) suggested a massive runaway star as a possible
progenitor of Kepler's SNR, since it was able to naturally account for
the presence of the CSM and the asymmetric structure of the remnant in
optical wavelength.  Recently, Blair et al.\ (2007) proposed that the
Kepler's SN was categorized as a Type-{\scshape I}a explosion in a
region with significant CSM, which was a small but growing class of
Type-{\scshape  I}a SNe named as Type-{\scshape I}a/IIn by Kotak et
al.\ (2004).  Due to the large expansion of Kepler's SNR, it is the
best target from which we can investigate the detailed preexisting
structures of the CSM for this rare kind of SNe. 

Here, we report proper-motions of the forward shocks on the overall
rims of Kepler's SNR, by using archival {\it Chandra} data.  We derive
the distance to the remnant, combining the proper-motions we measure
with the optically determined forward shock velocity (Sankrit et al.\ 
2005).  Also, we present evolutional states in various portions of the
remnant, which gives us critical information on structures of the
surrounding CSM.

\section{Observations}

Kepler's SNR was observed on three epochs in 2000 (PI: Holt, S.), 2004
(PI: Rudnick, L.), and 2006 (PI: Reynolds, S.).  We use the first
(ObsID.\ 116) and the last (ObsID.\ 6715) observation to measure
proper-motions of the forward shocks of this remnant.  These
observations were, respectively, done in June 30th 2000 and August
4--6th 2006, resulting in the time difference of 6.09 yr.  The entire
remnant was covered on the ACIS-S3 back-illuminated chip in both
observations.  We start our analysis from level 2 event files
processed with calibration data files in CALDB ver.\ 3.4.0 for ObsID.\
116, ver.\ 3.2.2 for ObsID.\ 6715.  We exclude the high-background
periods for data from ObsID.\ 116, whereas there seems no significant
background flares for the data from ObsID.\ 6715 so that we reject no
data from the level 2 event file for this data set.  The resulting net 
exposure times for ObsID.\ 116 and 6715 are 37.8\,ks and 159.1\,ks,
respectively. 

To measure proper-motions of the forward shocks, we use an energy range
of 1.0--8.0\,keV, i.e., we do not use an energy range below 1.0\,keV
where building up contaminants on the detector significantly reduce
count rates for the second-epoch observation.  Figure~\ref{fig:image}
{\it left} shows the second-epoch image of the entire Kepler's SNR in
the energy range of 1.0--8.0\,keV.  

\section{Analysis and Results}

We register the two images by aligning them on four point sources that
are obviously seen in the vicinity of the remnant.  We determine the
positions of the four sources in both observations, employing {\tt
  wavdetect} software included in CIAO ver.\ 3.4.  We find a systematic
offset between the two images; the sense of the difference is that
the first-epoch image is south and west of the second.  We calculate
the error-weighted mean offset of the four point sources to be 
0$^{\prime\prime}$.14 in right ascension and 0$^{\prime\prime}$.19
in declination.  Then, the first-epoch image is shifted in right
ascension by 0$^{\prime\prime}$.14 and in declination by
0$^{\prime\prime}$.19 with respect to the second-epoch image.  Once
the two images are registered, we do not need to consider the absolute 
astrometric uncertainty of 0$^{\prime\prime}$.6 reported by the 
{\it Chandra} calibration team as systematic errors in our analysis.
We take account of 0$^{\prime\prime}$.1, i.e., the relative
astrometric uncertainty reported by the {\it Chandra} calibration
team, as systematic errors in the following proper-motion
measurements.  

\begin{figure}
\includegraphics[angle=0,scale=0.45]{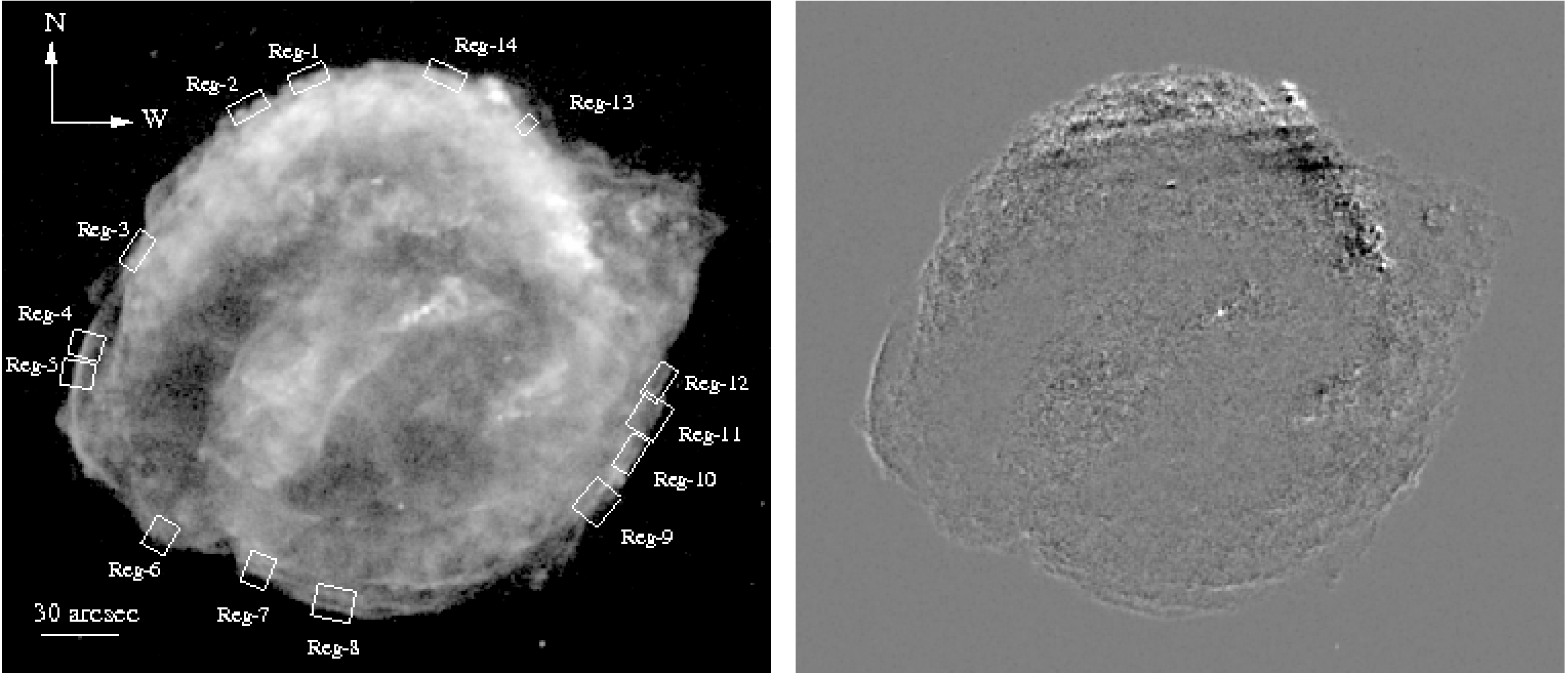}\hspace{1cm}
\caption{{\it Left}: {\it Chandra} 1.0--8.0\,keV band image obtained
  in the second-epoch observation.  The image is binned by
  0$^{\prime\prime}$.492 and has been smoothed by Gaussian kernel of
  $\sigma = 0^{\prime\prime}$.984.  The intensity scale is logarithmic.  
  We measure proper-motions of the forward shock in 14 regions
  indicated as rectangules (from Reg-1 to Reg-14).  {\it Right}:
  Linearly scaled difference (2006$-$2000) image between the two epochs.  
} 
\label{fig:image}
\end{figure}

Figure~\ref{fig:image} {\it right} shows a difference image between
two epochs after being registered and normalized to match the
count rates in the two epochs.  We clearly see a signature of the
expansion as positive emission (seen as white in the figure) with
negative emission (seen as black in the figure).  Note that the
horizontal stripes seen in the northern rim of the remnant are due
to bad columns on the ACIS-S3 chip.

\begin{figure}
\includegraphics[angle=0,scale=0.65]{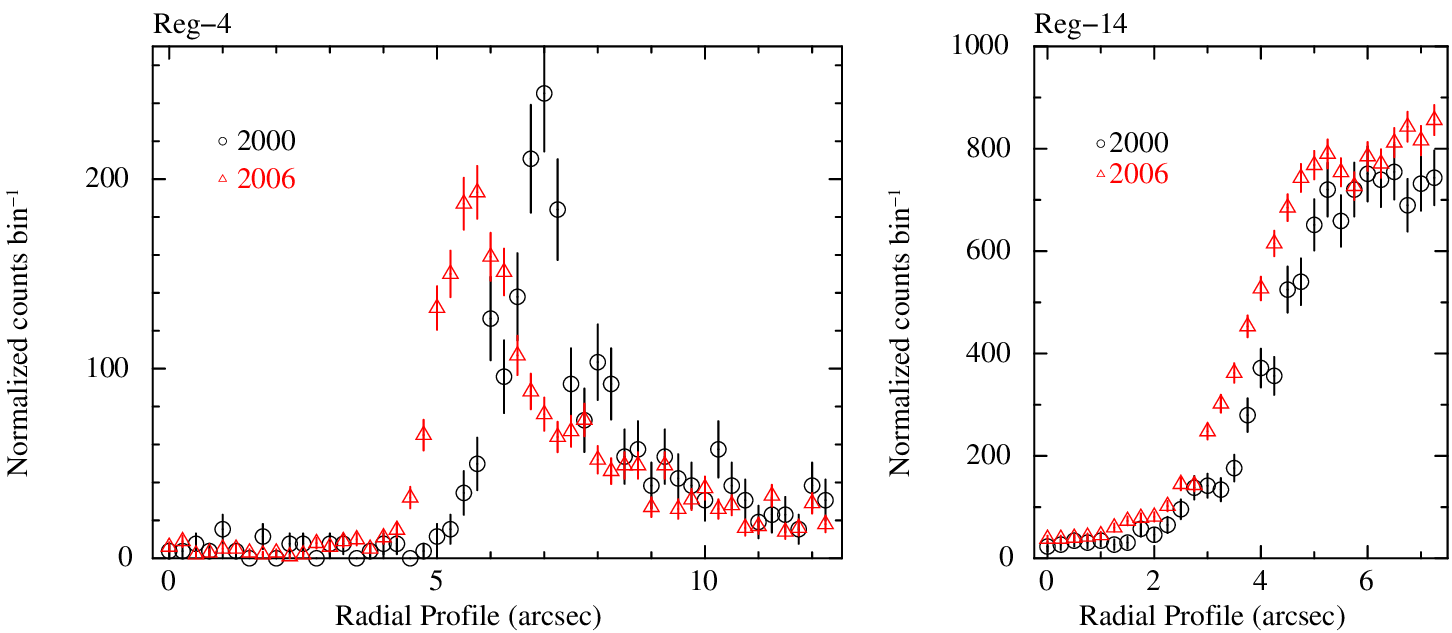}
\caption{{\it Left}: Radial profiles binned with a
  0$^{\prime\prime}$.25 scale derived for Reg-4.  Data points with circles and
  triangles represent the first- and the second-epoch observations,
  respectively.   The shock motion is in the left direction.  {\it
  Right}: Same as {\it left} but for Reg-14.}    
\label{fig:profiles}
\end{figure}

To measure proper-motions of the forward shocks of the remnant, we
focus on 14 rectangular regions at the edge of the X-ray image
(see, Fig.~\ref{fig:image} {\it left}).  We then project the image
into one dimension so that we generate one-dimensional profile along
perpendicular direction to the shock motion (hereafter, radial
profile).  Each bin of the radial profile is spaced by
0$^{\prime\prime}$.25.  We adjust the position angle for each
rectangular region as tangential as possible to the shock front in the
following way.  We generate radial profiles for position angle with
trial angles of 2$^\circ$ steps.  For each trial position angle, we
calculate 
\[I = \sum_{i} I_{i}^2,
\]
where $I_{i}$ is the intensity in each bin, $i$.  The position angle
with the largest value of $I$ represents the most tangential angle to the
shock front.  The 14 regions with the ``best'' position angles are
shown in Fig.~\ref{fig:image} {\it left}.  We plot example radial
profiles from Reg-4 and Reg-14 in Fig.~\ref{fig:profiles}, from which 
we can see apparent shifts between the two epochs.  To quantitatively
measure the shifts, the first-epoch profile is shifted in radius with
respect to the second-epoch profile, and $\chi^2$ is calculated from
the difference between the two profiles at each shift position (e.g.,
Katsuda et al.\ 2008).  In
this way, we obtain $\chi^2$ profiles as a function of shift
positions.  By applying a quadratic function for the $\chi^2$ profile,
we measure the best-shift position where the minimum $\chi^2$-value
($\chi^2_\mathrm{ min}$) occurs.  90\% statistical uncertainties on
the best-shift are calculated using a criteria of $\chi^2_\mathrm{min}
+ 2.7$.  Table.~\ref{tab:summary} summarizes the best-shift positions
between the two observations, $\chi^2_\mathrm{min}$ per degrees of
freedom, and proper motions for all the regions indicated in
Fig.~\ref{fig:image} {\it left}.  The azimuthal angle for each region, 
which is measured counterclockwisely from the radio center
at $\alpha = $ 17$^{\mathrm h}$30$^{\mathrm m}$41$^{\mathrm s}$.5,
$\delta = -21^{\circ}29^{\prime}23^{\prime\prime}$ (J2000; Matsui et  
al.\ 1984), is also listed in the table.

\begin{figure}
\includegraphics[angle=0,scale=0.65]{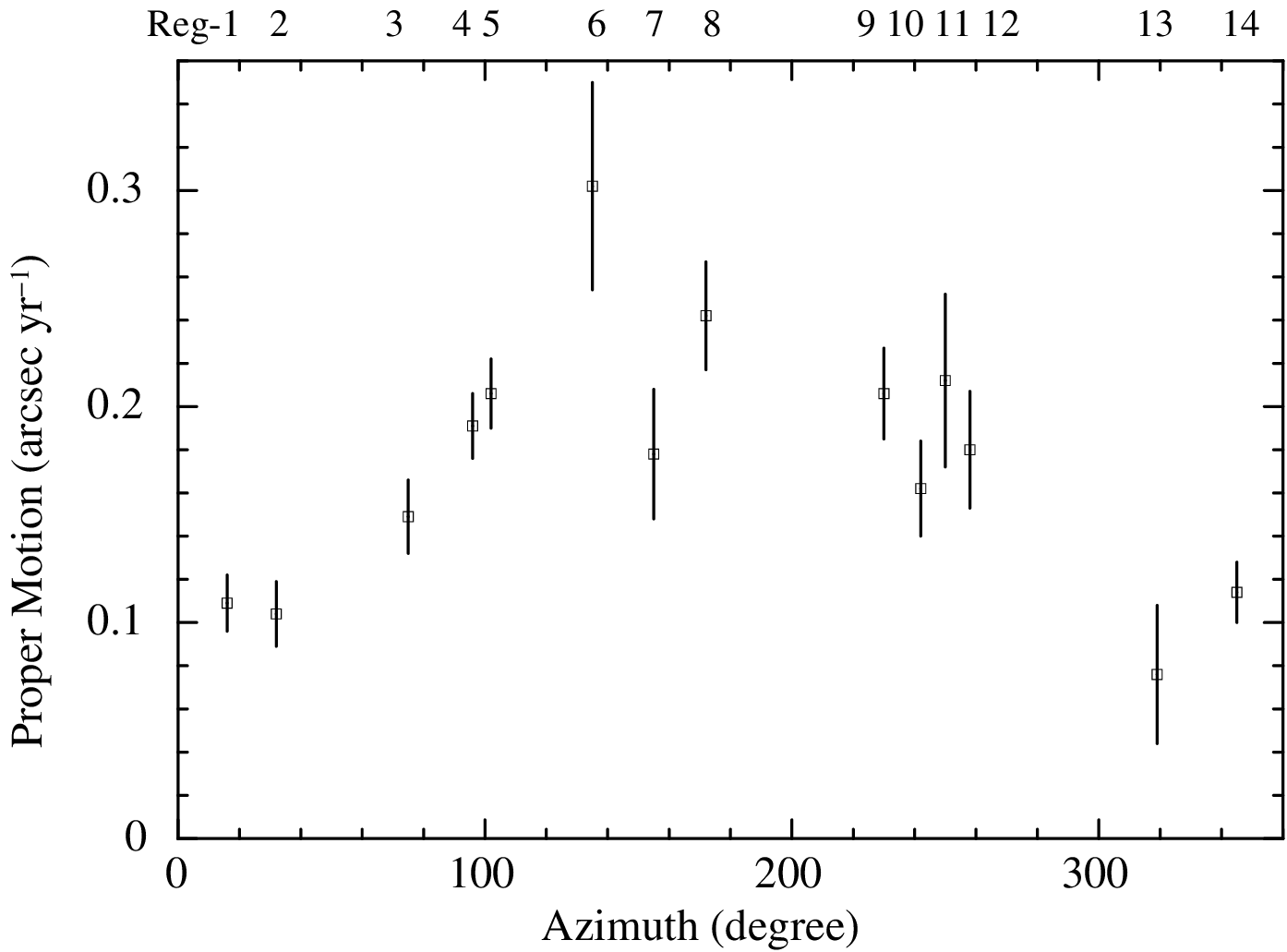}
\caption{Proper motions as a function of azimuthal angle (lower
  x-axis) and region number (upper x-axis).  Note that
  quated errors represent only statistical uncertainties.} 
\label{fig:azimuth}
\end{figure}

We find that the proper-motions vary from location to location.
Plotting the proper-motion derived in each region as a function of
azimuthal angle in Fig.~\ref{fig:azimuth}, we find a trend that the
proper-motions derived in the northern regions (i.e., Regs-1, 2, 13,
and 14) are slower than those derived in the rest of the regions.  

\section{Discussion}

\subsection{Distance to the Remnant}

We have measured proper-motions of the forward shocks on overall
rims of Kepler's SNR.  If we combine them with the shock
velocities, the distance to the remnant can be determined.  The
velocities of the forward shocks associated with Balmer-dominated
filaments, which were located at the northern rim and the central
portion of the remnant, were determined to be
1670--2800\,km\,sec$^{-1}$ (Fesen et al.\ 1989) or
1550--2000\,km\,sec$^{-1}$ (Blair et al.\ 1991) from their H$\alpha$  
emission line widths.  Under the assumption of little or no
temperature equilibration between electrons and protons, Sankrit et
al.\ (2005) determined the 
best-estimated shock velocity to be 1660$\pm$120\,km\,sec$^{-1}$.
Since the best-estimated shock velocity represents the average
velocity of the forward shocks with Balmer-dominated filaments seen in
the northern and the central portion of the remnant, we should combine it
with an error-weighted mean proper-motion measured in the northern
regions (i.e., Regs-1, 2, 13, 14).  Then, the best-estimated distance
to the remnant is determined to be 3.3
($v$/1660\,km\,sec$^{-1}$)($\mu$/0$^{\prime\prime}$.107
yr$^{-1}$)$^{-1}$ kpc.  We can estimate a distance range of
2.9--4.9\,kpc to the remnant, considering the variation of the
proper-motions derived in the four northern regions of 
0$^{\prime\prime}$.076--0$^{\prime\prime}$.11 yr$^{-1}$ as well as the
uncertainty of the shock velocity.  While the best-estimated distance
determined here is less than $\sim$5\,kpc (Reynoso \& Goss 1999), it
is well within the values previously measured based on the combination
of the proper-motion and the shock-velocity of the Balmer-dominated
filaments: from 2.9\,kpc (Blair et al.\ 1991) to 3.9\,kpc (Sankrit et
al.\ 2005).  

\subsection{Asymmetry of the Forward Shock Velocity}

We find velocity asymmetries of the forward shocks of Kepler's SNR:
shock velocities on the northern rim are 1.5--3 times slower than
those on the other rims.  If we assume the pressure equilibrium in the
remnant, the velocity contrast of 1.5--3 requires a density contrast
of $\sim$2--9.  Recently, Blair et al.\ (2007) estimated that the
preshock density in the north of the remnant is $\sim$4--9 times
higher than that in the south, based on brightness variations observed
between the northern and southern rims of the remnant at 24\,$\mu$m.
Therefore, the slower expansion observed on the northern rim than
those on the rest of the rims seems to be well explained by the
density contrast suggested from the observation at 24\,$\mu$m.   

\begin{figure}
\includegraphics[angle=0,scale=0.65]{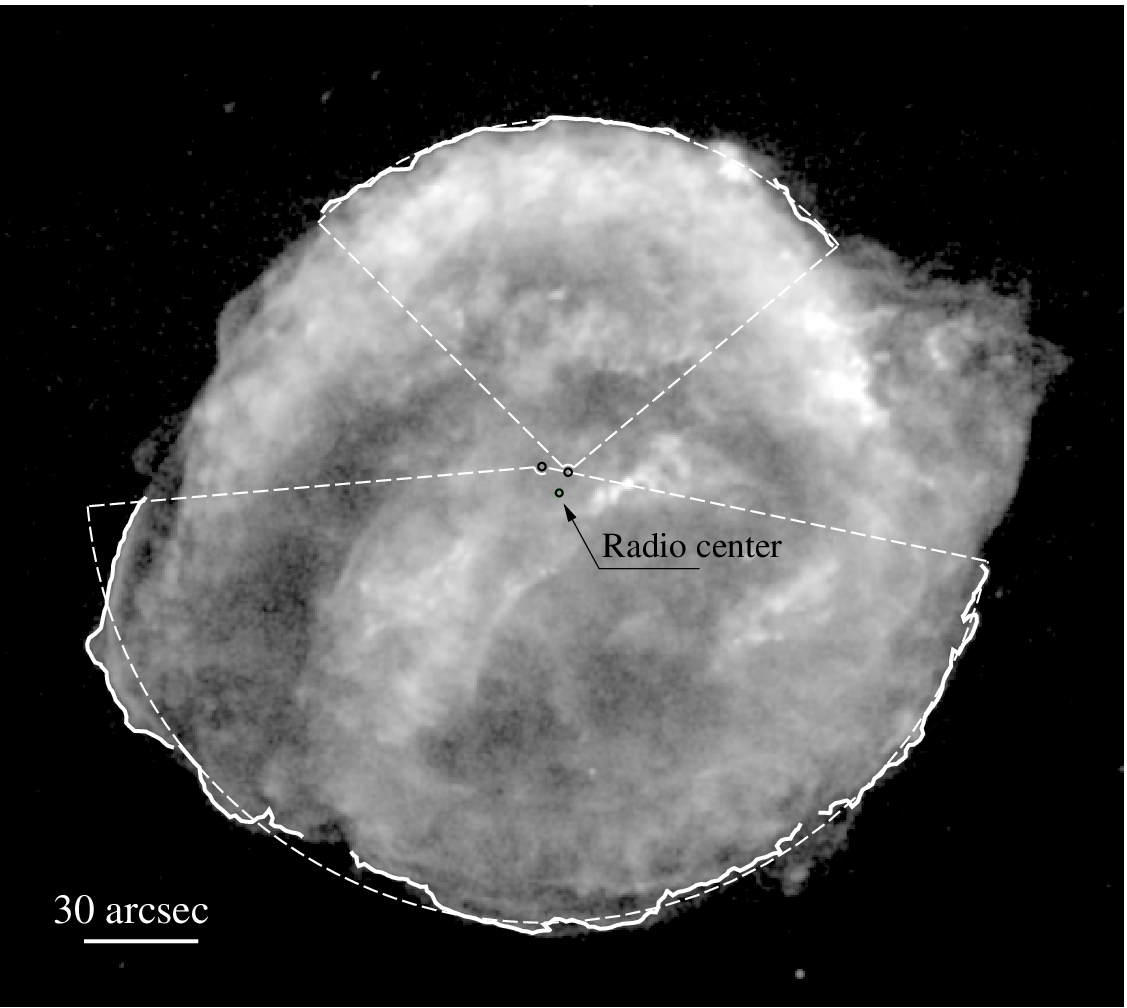}
\caption{Best-fit circles to represent the X-ray boundaries of the
  northern and southern halves of the remnant are indicated as white
  dashed pie superposed on the second-epoch {\it Chandra}
  0.3--8.0\,keV band image.  
  The radio center (Matsui et al.\ 1984) is also indicated as a small
  black circle.  White contours (5 counts per pixel) are the X-ray
  boundaries used to estimate the best-fit circles shown in this image. 
}    
\label{fig:center}
\end{figure}

It is interesting that Kepler's SNR appears quite round in spite of
the asymmetric velocities of the forward shocks.  One possibility to
explain this feature is that the forward shock encountered a dense gas
on the northern rim so recently that we can hardly see apparent
deceleration of the shock from morphological point of view.  
However, there have been no reports of such a very recent shock
deceleration so far.  (In addition, we can not obtain at least
strong evidence that the forward shock on the northwestern rim shows
recent significant deceleration; Sankrit et al.\ [2005] 
measured a proper-motion of a Balmer-dominated filament positionally
coincident with the forward shock in Reg-13 from observations
performed on two epochs in 1987 and 2003 to be
0$^{\prime\prime}$.089$\pm$0$^{\prime\prime}$.009 yr$^{-1}$ that is
consistent with our estimation of the proper-motion of
0$^{\prime\prime}$.076 [$\pm$0$^{\prime\prime}$.032
$\pm$0$^{\prime\prime}$.016] yr$^{-1}$ determined in between 2000 and
2006).  We propose another possibility that the expansion center of
the remnant is located at a relatively northern position compared to
the geometric center 
determined by the entire remnant.  In fact, the outer edges of the
X-ray structure seem to be outlined by two (ideally concentric)
circles whose centers are located at the north of the radio center: one with
a relatively small radius outlines the northern-half edge of the
remnant, whereas the other with a relatively large radius outlines
the southern-half edge.  We estimate the best-fit circle for either
the northern-half edge or the southern-half edge independently.  Here, 
we define the northern half in a range of azimuthal angles from
$-$50$^\circ$ (or 310$^\circ$) to 45$^\circ$, whereas the
southern half in a range of azimuthal angles from 95$^\circ$ to
258$^\circ$, such that the northern half covers the regions with
proper-motions of $\sim$0$^{\prime\prime}$.1 yr$^{-1}$, 
whereas the southern half covers the other regions with proper-motions
of above $\sim$0$^{\prime\prime}$.2 yr$^{-1}$.  We define the edge
of the X-ray extent (hereafter, X-ray boundary) as contours of
5 counts per 0$^{\prime\prime}$.492-sided pixel in the second-epoch
0.3--8.0\,keV band image.  We ignore X-ray boundaries showing
apparent deviations from circular curvatures, in order not to yield
misleading geometric centers and radii.  The X-ray
boundaries used to estimate the best-fit circles are drawn by
white curves (contours) in Fig.~\ref{fig:center}.  Assuming 
the geometric center, we calculate the distance, $R_i$, between
each pixel, $i$, on the X-ray boundaries and the geometric center.
For various trial geometric centers, we calculate 
\[K = \sum_{i} (R - R_{i})^2,
\]
where $R$ is a variable parameter representing a radius.  We can
derive the best-fit radius, $R$, as well as the geometric center
at the minimum of the $K$-value.  The best-fit circles representing the
X-ray boundaries for the northern and southern halves are centered at
$\alpha = $ 17$^{\mathrm h}$30$^{\mathrm m}$41$^{\mathrm s}$.4, $\delta = 
-21^{\circ}29^{\prime}19^{\prime\prime}$ (J2000) with a radius of
93$^{\prime\prime}$, and at $\alpha = $ 17$^{\mathrm h}$30$^{\mathrm
  m}$41$^{\mathrm s}$.9, $\delta =
-21^{\circ}29^{\prime}16^{\prime\prime}$ (J2000) 
with a radius of 120$^{\prime\prime}$, respectively.  Parts of these
two circles are indicated in Fig.~\ref{fig:center} as white dashed
pie. We see that both geometric centers estimated are
shifted by about 5$^{\prime\prime}$ toward the north of the remnant
from the radio center (see, Fig.~\ref{fig:center}), which
results in a smaller radius for the circle representing the
northern-half X-ray boundary than that for the southern-half X-ray
boundary; the ratio of the radius is derived to be about 3:4.  
Although we cannot derive uncertainties on the center positions and
radii in the least square method used above, we might estimate
uncertainties on these parameters from the $\chi^2$ method by
introducing fake errors on the data ($R_i$).  We introduce the fake 
errors of 1\,\% for the northern half and 3.5\,\% for the southern
half, respectively, so that we can derive the reduced $\chi^2$-value 
of $\sim$1.  The 90\% uncertainties, i.e., $\chi^2 <
\chi^2_\mathrm{min} + 6.25 $ as appropriate for three interesting
parameters, are estimated to be $\pm$0$^{\prime\prime}$.5 or
$\pm$1$^{\prime\prime}$.5 (in right ascension), and
$\pm$1$^{\prime\prime}$.5 or $\pm$2$^{\prime\prime}$ (in declination)
for the northern or southern halves, respectively.  Therefore, 
the derived offset of $\sim$5$^{\prime\prime}$ from
the radio center seems to be significant, although we cannot strongly
state the significance without knowing the uncertainty of the radio center.
Hydrodynamic simulations of a remnant expanding into a medium with a
density gradient (Dohm-Palmer \& Jones 1996) show that the shock
outline remains roughly circular, while the center of the best-fit
curvature can move away from the true explosion location by as much as
10 -- 15\% of the remnant radius.  The offsets between geometric
centers estimated here and the radio center (Matsui et al.\ 1984) are
roughly 5\% of the radius.  Therefore, from theoretical point of view,
our newly determined geometric centers can be possible true explosion
locations of Kepler's SNR.  We believe that the true explosion point
is around the geometric centers estimated here.

\begin{figure}
\includegraphics[angle=0,scale=0.65]{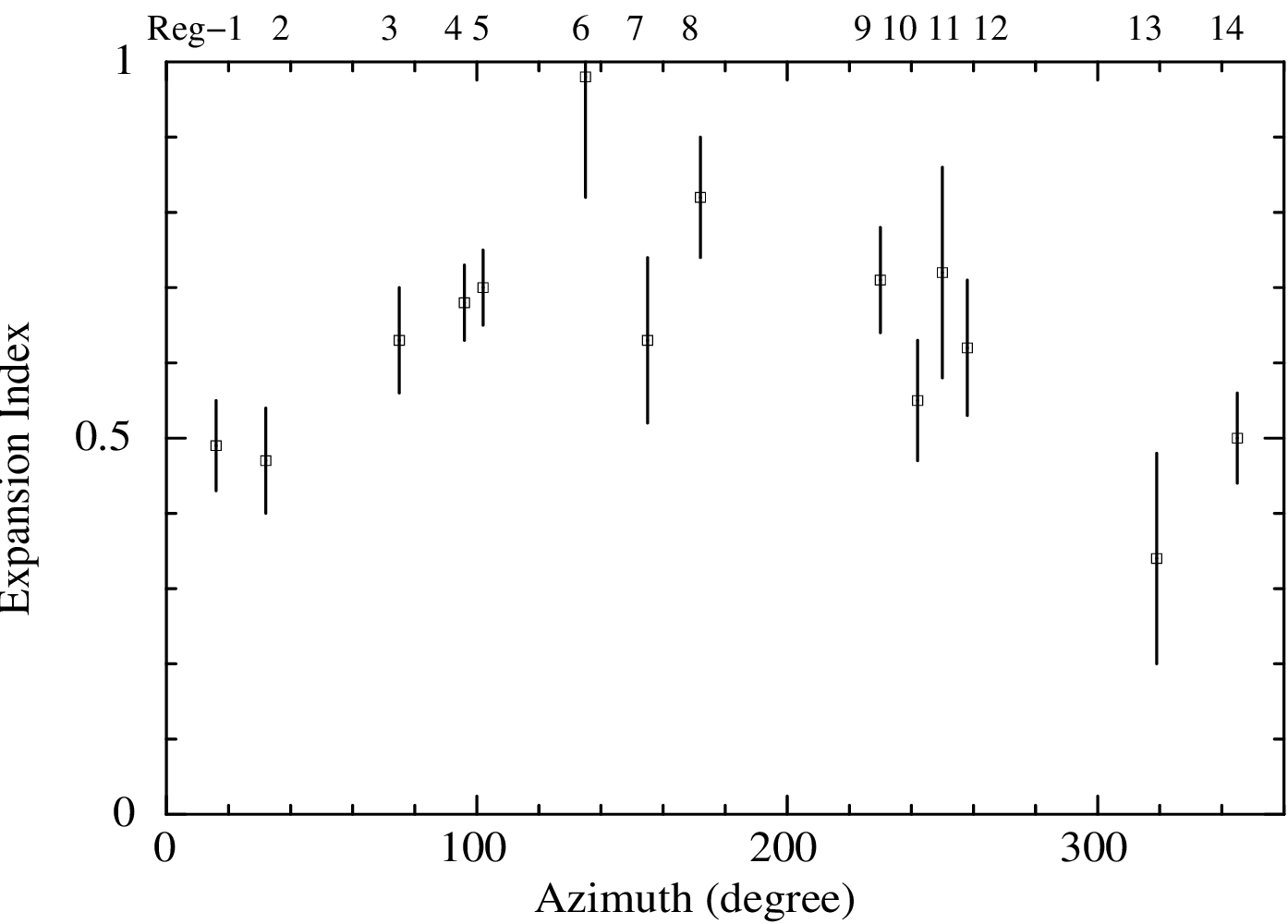}
\caption{Same as Fig.~\ref{fig:azimuth} but for expansion indices.} 
\label{fig:azimuth2}
\end{figure}

We consider the expansion center of Kepler's SNR at $\alpha =
$17$^{\mathrm h}$30$^{\mathrm m}$41$^{\mathrm s}$.6, $\delta = 
-21^{\circ}29^{\prime}17^{\prime\prime}$ (J2000) that is the center
point between the two center points of the circles representing the
northern-half and southern-half X-ray boundaries.  Then, the expansion
rate can be calculated by dividing the
proper-motion value in each region by the distance between the shock
front and the expansion center.  Combined with the age of the remnant
of 400 yr, the expansion index, $m$, is also calculated, where the
expansion of the forward shock of SNRs can be expressed as $R
\varpropto t^m$ (i.e., the remnant's radius is assumed to evolve as a
power law with age; see, e.g., Woltjer 1972).  Table.~\ref{tab:param}
summarizes these parameters and Fig.~\ref{fig:azimuth2} shows the
expansion indices as a function of azimuthal angle.  SNRs cease
being in pure free expansion ($m=1$) after a few years, as they interact with
circumstellar or interstellar gas.  Then, a reverse shock into the
ejecta is formed, and remnants evolve in what is usually called an
ejecta-driven phase.  In spherical symmetry, if both the ejecta and the CSM
have power-law density profiles characterized by $\rho \varpropto
r^{-n}$ and $\rho \varpropto r^{-s}$, respectively, the evolution becomes
self-similar and is given by $R \varpropto t^{(n-3)/(n-s)}$, i.e., the
expansion index, $m$, can be written as $(n-3)/(n-s)$ (e.g., Chevalier
1982).  Here, the value of $s$ is 0 for a uniform CSM, and is 2 for a
constant wind velocity from the progenitor star.   
In general, the $s$-value is expected to be 2 for core-collapse
SNe, since massive stars which results in core-collapse SNe produce
stellar wind before SN explosions, while both $s = 0$ and $s = 2$ can 
happen for Type-{\scshape I}a SNe as mentioned in Section 1.  As for
$n$-values, $9 \lesssim n \lesssim 12$ is expected for a core-collapse
origin (e.g., Chevalier 1992), while $5 \lesssim n \lesssim 7$ for a
Type-Ia origin (e.g., Chevalier 1982).  Therefore, for core-collapse
SNe, the value of $m$ is expected to be greater than 6/7 (=0.86).  On
the other hand, for Type-{\scshape I}a SNe, the value of $m$ ranges
from 0.4 to 0.8.  Our measured $m$-values at all regions except for
regions 6 and 13 range from 0.47 to 0.82, which is consistent with
that expected from Type-{\scshape I}a SNe.  We find significant
variations of $m$-values in this 
remnant, which suggests complicated CSM structures around the
remnant.  It should be remarked that the northern rim shows
slower expansions than the southern rim does.  The north-south density 
asymmetry of the CSM suggested from variations of surface brightness
results in the variation of $m$-values between the northern rim and
the southern rim.

\begin{figure}
\includegraphics[angle=0,scale=0.65]{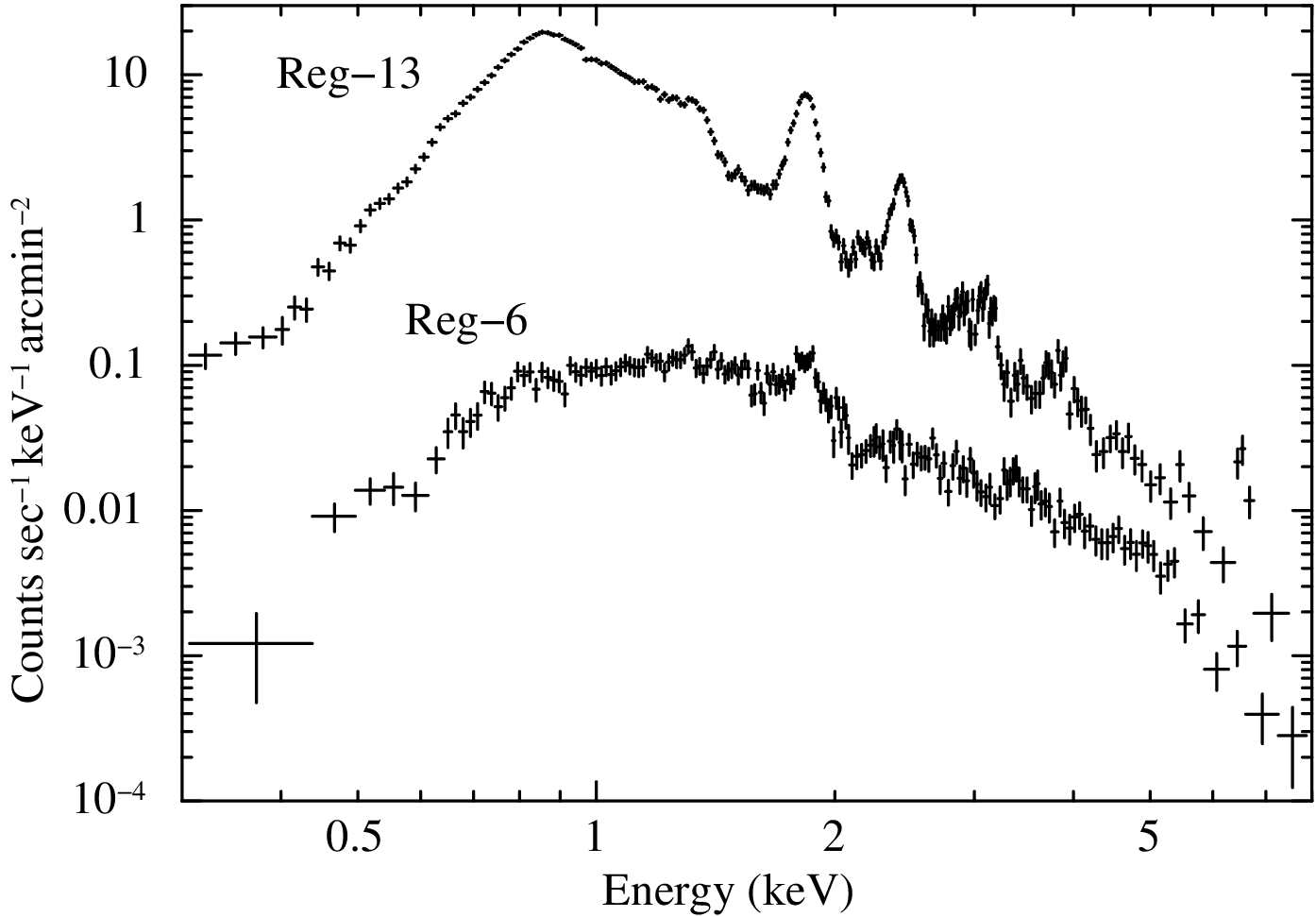}
\caption{Spectra extracted from the southeastern rim around Reg-6
  and the northwestern rim around Reg-13.}
\label{fig:spec}
\end{figure}

It is worth noting how the velocity difference of the forward shock
affects spectral features.  As already noticed by {\it XMM-Newton}
(Cassam-Chena$\ddot{\mathrm i}$ et al.\ 2004) and {\it Chandra}
(Reynolds et al.\ 2007) observations, a spectrum extracted around
Reg-6 on the southeastern rim is non-thermal (synchrotron) in origin,
whereas a spectrum around Reg-13 on the northwestern rim is thermal in
origin.  We can clearly see the spectral difference in
Fig.~\ref{fig:spec}.  This is considered as a result that the faster
shock (4700\,km\,sec$^{-1}$ at a distance of 3.3\,kpc) as well as the
lower ambient density suggested by lower surface brightness in Reg-6
produces the synchrotron emission more efficiently than the slower
(1200\,km\,sec$^{-1}$ at a distance of 3.3\,kpc) shock as well as
higher ambient density in Reg-13 does. 

We should note previous X-ray expansion measurements based on {\it
Einstein} and {\it ROSAT} observations performed by Hughes (1999).
The mean expansion rates in the entire remnant was derived to be
$\sim$0.24\% yr$^{-1}$.  This value is larger than that
estimated in our analysis; only one region (Reg-6) shows such a rapid
expansion.  Furthermore, he measured expansion rates as a function of
azimuthal angle, by comparing long radial profiles from the geometric
center to the X-ray boundaries between two {\it ROSAT} HRI
observations.  The expansion rates measured in the northern portion of
the remnant did not show lower values than those in the other
portions.  This also conflicts to our results.  These discrepancies
might come from the different method between Hughes (1999) and this
work.  Hughes (1999) measured radially averaged proper-motions of the
remnant, on the other hand, we measure proper-motions for the very
edge of the remnant, i.e., the forward shocks themselves.  Therefore,
the faster expansion rates derived in Hughes (1999) than those in this
work suggests that plasmas in the inner remnant show larger expansion
rates than those of the forward shocks.  Such a situation might
indicate recent rapid deceleration of the forward shocks.  However, as
mentioned above, we have not yet observed (at least strong) such 
indications so far.  Further detailed proper-motion measurements in
the inner remnant are strongly required to reveal the reason of the
discrepancies between the expansion rate by Hughes (1999) and that in
this work.  

\section{Conclusions}

We have measured proper-motions of the forward shocks on the overall 
rims of Kepler's SNR, using the archival {\it Chandra} data.

The expansion indices measured at various parts of the rim supports
the Type-{\scshape I}a SN which has the situation that the ejecta and
the CSM, respectively, have power-law density profiles with indices
of 5--7 and 0--2 rather than the core-collapse SN. 

We find that the shock velocities are
asymmetric: the shock velocities on the northern rim are 1.5--3 times
slower than those on the rest of the remnant.  We attribute this
asymmetry to the density inhomogeneities of the CSM surrounding the
remnant.  The shape of the X-ray boundary of the remnant as well as
the inhomogeneous CSM structures lead us to consider that the
expansion center is located at $\sim$5$^{\prime\prime}$ offset in the
north of the radio center.  

\acknowledgments

This work is partly supported by a Grant-in-Aid for Scientific Research
by the Ministry of Education, Culture, Sports, Science and Technology
(16002004).  S.K.\ is supported by JSPS Research Fellowship for Young
Scientists.


\begin{deluxetable}{ccccc}
\tabletypesize{\scriptsize}

\tablecaption{Summary of Proper-Motion Measurements}
\tablewidth{0pt}
\tablehead{
\colhead{Region} &\colhead{Azimuth (deg)} &\colhead{$\chi^2_\mathrm{
    min}$/d.o.f.} &\colhead{Shift between the two epochs (arcsec)}  &
\colhead{Proper motion (arcsec yr$^{-1}$)}  
}
\startdata
Reg-1 & 16 & 1.68 & 0.63 ($\pm$0.09 $\pm$0.10) & 0.109 ($\pm$0.013 $\pm$0.016)\\
Reg-2 & 32 & 0.54 & 0.66 ($\pm$0.08 $\pm$0.10) & 0.104 ($\pm$0.015 $\pm$0.016)\\
Reg-3 & 75 & 0.85 & 0.91 ($\pm$0.11 $\pm$0.10) & 0.149 ($\pm$0.017 $\pm$0.016)\\
Reg-4 & 96 & 1.47 & 1.16 ($\pm$0.09 $\pm$0.10) & 0.191 ($\pm$0.015 $\pm$0.016)\\
Reg-5 & 102 & 2.27 & 1.26 ($\pm$0.10 $\pm$0.10)& 0.206 ($\pm$0.016 $\pm$0.016)\\
Reg-6 & 135 & 1.56 & 1.84 ($\pm$0.29 $\pm$0.10)& 0.302 ($\pm$0.048 $\pm$0.016)\\
Reg-7 & 155 & 1.71 & 1.08 ($\pm$0.18 $\pm$0.10)& 0.178 ($\pm$0.030 $\pm$0.016)\\
Reg-8 & 172 & 1.39 & 1.48 ($\pm$0.16 $\pm$0.10)& 0.242 ($\pm$0.025 $\pm$0.016)\\
Reg-9 & 230 & 1.80 & 1.25 ($\pm$0.13 $\pm$0.10)& 0.206 ($\pm$0.021 $\pm$0.016)\\
Reg-10 & 242 & 1.00 & 0.99 ($\pm$0.13 $\pm$0.10)& 0.162 ($\pm$0.022 $\pm$0.016)\\
Reg-11 & 250 & 1.25 & 1.29 ($\pm$0.24 $\pm$0.10)& 0.212 ($\pm$0.040 $\pm$0.016)\\
Reg-12 & 258 & 1.06 & 1.10 ($\pm$0.17 $\pm$0.10)& 0.180 ($\pm$0.027 $\pm$0.016)\\
Reg-13 & 319 & 0.89 & 0.46 ($\pm$0.19 $\pm$0.10)& 0.076 ($\pm$0.032 $\pm$0.016)\\
Reg-14 & 345 & 1.06& 0.70 ($\pm$0.08 $\pm$0.10)& 0.114 ($\pm$0.014 $\pm$0.016)\\

\enddata

\tablecomments{First-term errors represent 90\% statistical
  uncertainties and second-terms errors represent systematic
  uncertainties. } 
\label{tab:summary}
\end{deluxetable}

\begin{deluxetable}{ccccc}
\tabletypesize{\scriptsize}

\tablecaption{Summary of Expansion Rates and Expansion Indices}
\tablewidth{0pt}
\tablehead{
\colhead{Region} &\colhead{Azimuth (deg)} &\colhead{Exapnsion Rate (\%)}
&\colhead{X-Ray Expansion Index}  & \colhead{Radio
  Expansion Index$^{\mathrm a}$}  
}
\startdata
Reg-1 & 16 & 0.12 ($\pm$0.01 $\pm$0.02)& 0.49 ($\pm$0.06 $\pm$0.07)& 0.45\\
Reg-2 & 32 & 0.12 ($\pm$0.02 $\pm$0.02)& 0.47 ($\pm$0.07 $\pm$0.07)& 0.45\\
Reg-3 & 75 & 0.16 ($\pm$0.02 $\pm$0.02)& 0.63 ($\pm$0.07 $\pm$0.07)& 0.65\\
Reg-4 & 96 & 0.17 ($\pm$0.01 $\pm$0.01)& 0.68 ($\pm$0.05 $\pm$0.06)& ---\\
Reg-5 & 102 & 0.17 ($\pm$0.01 $\pm$0.01)& 0.70 ($\pm$0.05 $\pm$0.05)& ---\\
Reg-6 & 135 & 0.25 ($\pm$0.04 $\pm$0.01)& 0.98 ($\pm$0.16 $\pm$0.05)& ---\\
Reg-7 & 155 & 0.16 ($\pm$0.03 $\pm$0.01)& 0.63 ($\pm$0.11 $\pm$0.06)& ---\\
Reg-8 & 172 & 0.20 ($\pm$0.02 $\pm$0.01)& 0.82 ($\pm$0.08 $\pm$0.05)& ---\\
Reg-9 & 230 & 0.18 ($\pm$0.02 $\pm$0.01)& 0.71 ($\pm$0.07 $\pm$0.06)& ---\\
Reg-10 & 242 & 0.14 ($\pm$0.02 $\pm$0.01)& 0.55 ($\pm$0.08 $\pm$0.06)& 0.55\\
Reg-11 & 250 & 0.18 ($\pm$0.03 $\pm$0.01)& 0.72 ($\pm$0.14 $\pm$0.05)& 0.55\\
Reg-12 & 258 & 0.16 ($\pm$0.02 $\pm$0.01)& 0.62 ($\pm$0.09 $\pm$0.06)& 0.55\\
Reg-13 & 319 & 0.08 ($\pm$0.04 $\pm$0.02)& 0.34 ($\pm$0.14 $\pm$0.07)& 0.35\\
Reg-14 & 345 & 0.12 ($\pm$0.02 $\pm$0.02)& 0.50 ($\pm$0.06 $\pm$0.07)& 0.35\\

\enddata
\tablecomments{First-term errors represent 90\% statistical uncertainties and
  second-term errors represent systematic uncertainties.  
  $^{\mathrm a}$Dickel et al.\ (1988) without particulary
  uncertain data.} 
\label{tab:param}
\end{deluxetable}

\end{document}